\documentclass[preprint,12pt]{elsarticle}
\usepackage{lineno}
\usepackage{amsmath}
\usepackage{multirow}

\journal{Journal of Instrumentation}

\biboptions{sort&compress}

\begin{document}

\begin{frontmatter}

\title{Characterization of VUV4 SiPM for Liquid Argon Detector}

\renewcommand{\thefootnote}{\fnsymbol{footnote}}
\author{
L.~Wang$^{a}$,
M.Y.~Guan$^{a,b,c}$,
H.J.~Qin$^{d}$,
C.~Guo$^{a,b,c}\footnote{Corresponding author. Tel:~+86-01088236256. E-mail address: guocong@ihep.ac.cn (C.~Guo).}$,
X.L.~Sun$^{a,b,c}\footnote{Corresponding author. Tel:~+86-01088236069. E-mail address: sunxl@ihep.ac.cn (X.L.~Sun)}$
C.G.~Yang$^{a,b,c}$,
Q.~Zhao$^{b,c}$,
J.C.~Liu$^{b,c}$,
P.~Zhang$^{a,b,c}$,
Y.P.~Zhang$^{a,b,c}$,
W.X.~Xiong$^{b,c}$,
Y.T~Wei$^{b,c}$,
Y.Y.~Gan$^{b,c}$,
J.J.~Li$^{e}$,
}
\address{
${^a}$ State Key Laboratory of Particle Detection and Electronics, Beijing, China\\
${^b}$ Experimental Physics Division, Institute of High Energy Physics, Beijing, China\\
${^c}$ School of Physics, University of Chinese Academy of Science, Beijing, China\\
${^d}$ School of Internet of Things Engineering, JiangNan University, Wuxi, China\\
${^e}$ School of Nuclear Science and Engineering, North China Electric Power University, Beijing, China
}

\begin{abstract}
Particle detectors based on liquid argon are now recognised as an attractive technology for dark matter direct detection and coherent elastic neutrino-nucleus scattering measurement. A program using a dual-phase liquid argon detector with a fiducial mass of 200~kg to detect coherent elastic neutrino-nucleus scattering at Taishan Nuclear Power Plant has been proposed. SiPMs will be used as the photon sensor because of their high radio-purity and high photon detection efficiency. S13370-6050CN SiPM, made by Hamamatsu, is a candidate for the detector. In this paper, the characterisation of S13370-6050CN SiPM, including the cross talk and after pulse probabilities at liquid argon temperature and the temperature dependence of breakdown voltage, dark counting rate and relative quantum efficiency were presented.
\end{abstract}


\end{frontmatter}


\section{Introduction}
Coherent elastic neutrino-nucleus scattering (CE$\nu$NS), which was first theorised by Freedman in 1974~\cite{PRD1974}, is the dominant process for neutrinos with energies less than 100~MeV. The COHERENT collaboration achieved the first measurement of CE$\nu$NS by using a 14.6~kg CsI(Na) crystal detector to detect the neutrinos from the spallation neutron source at Oak Ridge National Laboratory (ORNL) in 2017~\cite{SCI2017}. Recently, they reported the first detection of CE$\nu$NS on argon using the CENNS-10 liquid argon detector~\cite{arxiv2020}. A dual-phase liquid argon time projection chamber (TPC) with 200~kg fiducial mass was proposed to measure $\bar{\nu_e}$-Ar CE$\nu$NS process in China~\cite{arxiv202012}. The detector is expected to be adjacent to the JUNO-TAO experiment~\cite{TAO}, which is about 35~m from a reactor core of Taishan Nuclear Power Plant. Since the energy of the recoil nucleus is concentrated in the sub-keV region and the event rate is very low, single photon detection ability and high detector radio-purity, including the target and detector components, are the basic detector requirements.

Silicon photo-multipliers (SiPM), which were originally developed in Russia in the mid-1980s~\cite{NIMA2006}, underwent rapid development in the past few years and are expected to be a possible replacement of conventional photo-multiplier tubes (PMT). Compared with PMTs, SiPMs have lower radioactive background and higher photon quantum efficiency, which make them very promising as the photon sensors for low background experiments. However, the gain of SiPM is usually $\sim$10$^6$, thus a suitable amplifier must be used to detect single photon. A set of SiPM and preamplifier which can be used at liquid argon temperature is crucial for our experiment.

The S13770-6050CN~\cite{datasheet} SiPM examined in this paper is specially developed for cryogenic experiments by Hamamatsu and has been widely used in liquid xenon (LXe) detectors~\cite{LXe1,LXe2}. However, liquid argon (LAr) detectors require a lower temperature of 87~K and its performance has not been studied in detail.

The main purpose of this work is to study the the possible use of the S13370-6050CN SiPM in our liquid argon detector. While our experiment is in the R\&D phase, we consider DarkSide-20~\cite{DSyellowbook} to be a useful term of comparison. According to Ref.~\cite{DSyellowbook}, the SiPM used in a dual-phase liquid argon detector needs to meet at least the following two criteria:

A. The dark counting rate (DCR) should be less than 0.1~Hz/mm$^2$ to keep pulse shape discrimination effective;

B. The total correlated noise probability, namely the sum of direct crosstalk (DiCT), delayed crosstalk (DeCT) and after pulse(AP), should be less than 60\% for energy reconstruction of the S2 signal~\cite{DSyellowbook}.

In addition to the measurement of DCR and correlated noises, the changes of break down voltage and relative quantum efficiency (QE) from 87~K up to room temperatures are also introduced in this paper.

\section{Experimental Setup}
\label{sec.setup}

\begin{figure}[htb]
\centering
\includegraphics[height=5.5cm]{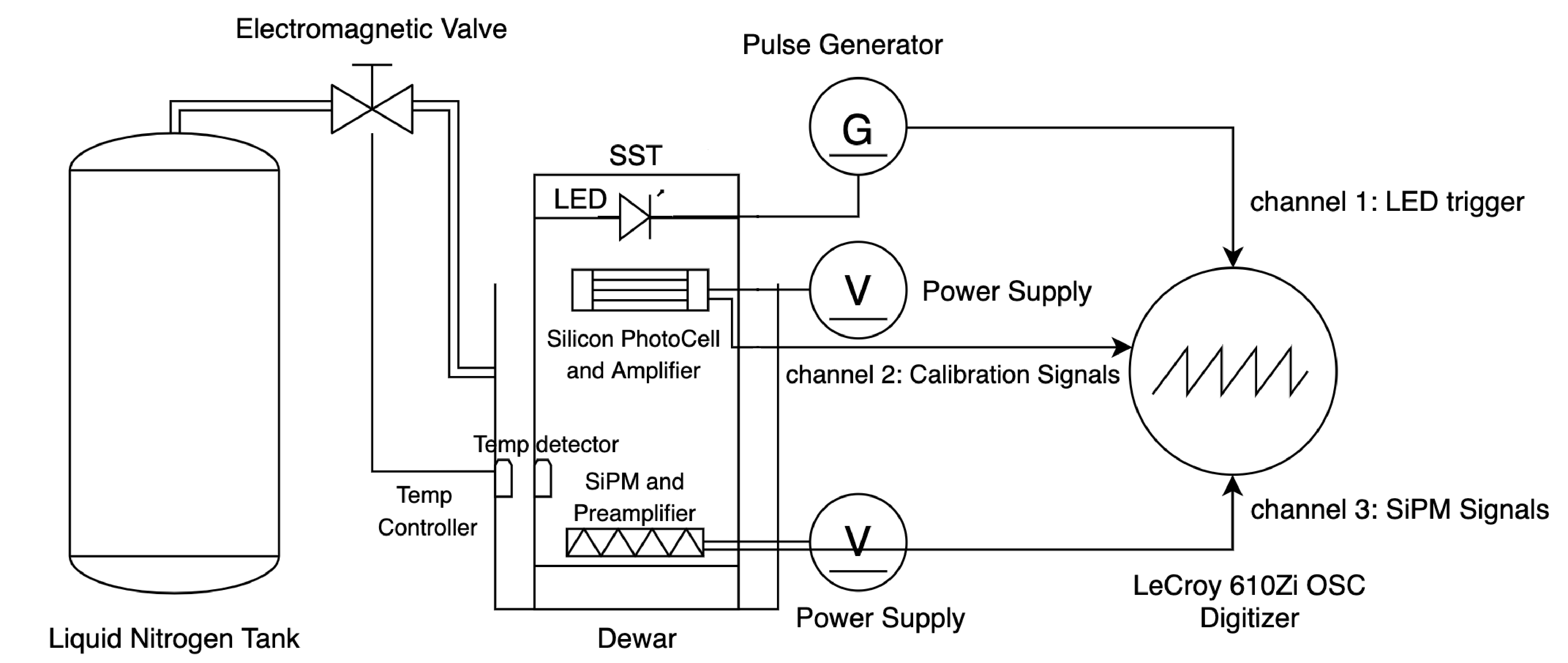}
\caption{Scheme of the experimental setup. }
\label{fig.setup}
\end{figure}

\begin{figure}[htb]
\centering
\includegraphics[width=5.5cm]{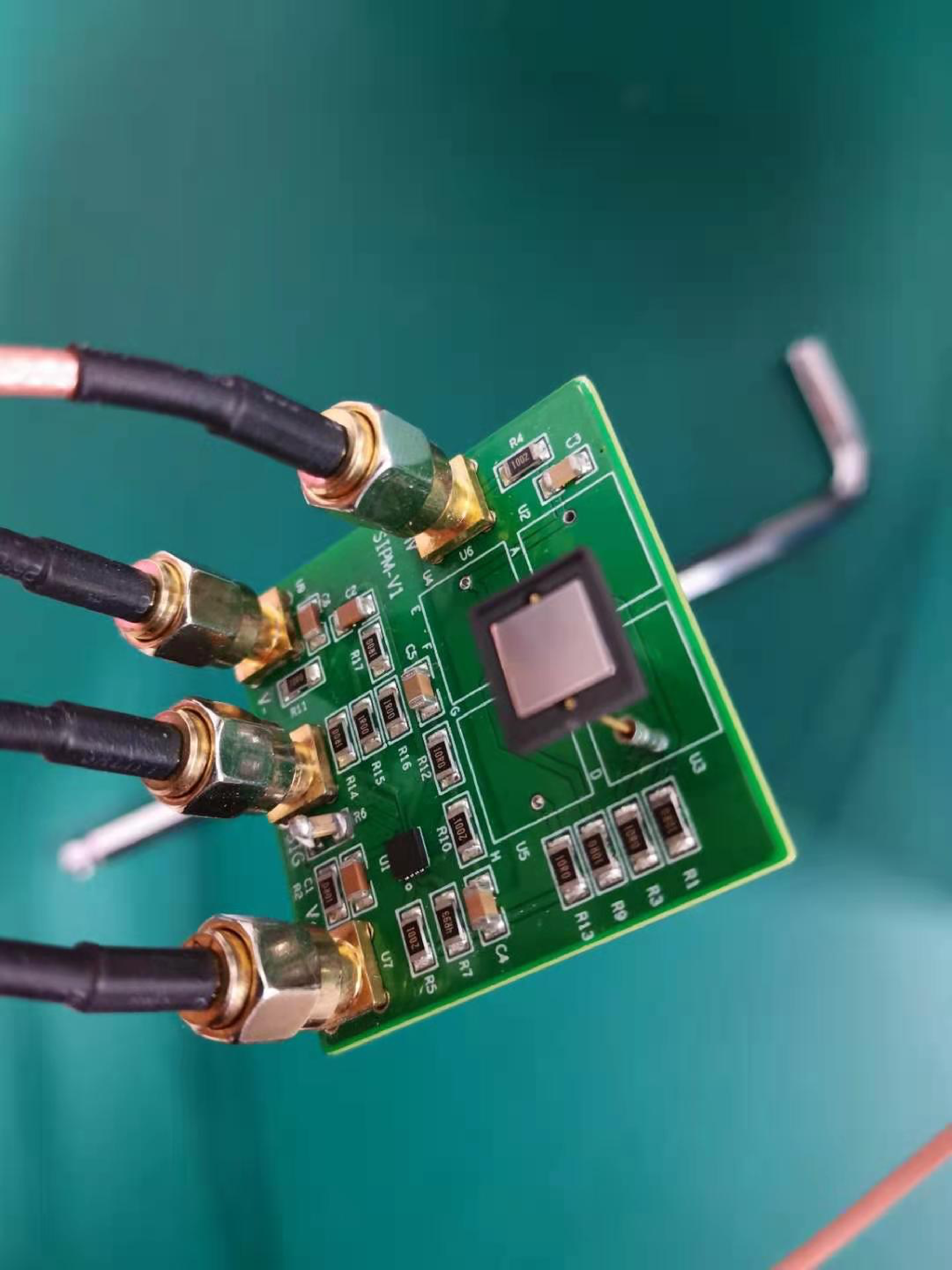}
\includegraphics[width=5.5cm]{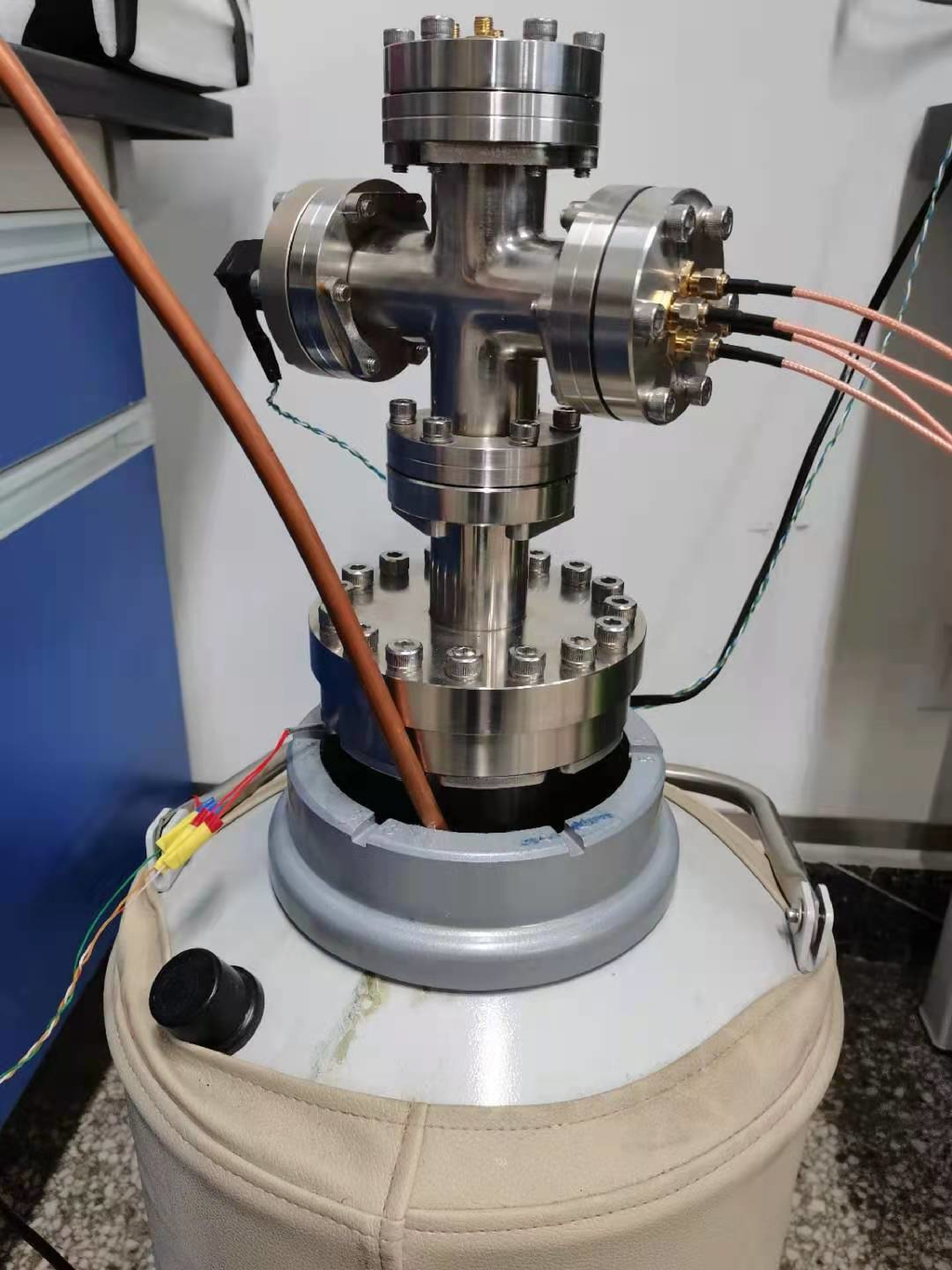}
\caption{Left: A 6 mm*6 mm SiPM connected with electronics. Right:A stainless steel (SST) chamber which is put in a dewar. }
\label{fig.sipm}
\end{figure}

The scheme of the experimental setup is shown in Fig.~\ref{fig.setup}. A 6~mm*6~mm SiPM (Fig.~\ref{fig.sipm} left-side) chip is placed in a stainless steel (SST) chamber which is put in a dewar (Fig.~\ref{fig.sipm} right-side). A liquid nitrogen tank is standing on the ground nearby for cooling the system by a copper tube. Two Pt100 temperature sensors have been used in this system as temperature detectors. One is located inside the SST chamber to monitor the temperature of the SiPM in real time, and the other is located in the inter-layer between the dewar and the SST chamber as the input signal of the temperature control system. It allows to intermittently inject liquid nitrogen by controlling the switch of the electromagnetic valve, so as to control the temperature. The accuracy of the temperature control system is $\pm$1~K and details can be found in Ref~\cite{NIMA980}. An LED with a peak wavelength of 425~nm is placed in the top of the SST chamber to be used as a light source. A silicon photocell (model LONGXINDA LXD44MQ) is placed next to the LED to monitor its stability~\cite{Photocell}. The LED is driven by a pulse generator (Tektronix, AFG31102, 100M, 2CH). During the whole experiment, the LED and the silicon photocell rest at room temperature. Before being sent into the oscilloscope (LeCroy 610Zi, 250MHz sampling frequency) for pulses recording, signals from the SiPM are amplified by a LMH6629 preamplifier which worked at the same temperature as the SiPM and signals from the silicon photocell are amplified by a warm amplifier (model AD825 LF353). Two RAGOL DP831A DC power supplies provide $\pm$5~V working power for two amplifiers separately and a DH 1765-4 DC power supply (1~mV accuracy) serves as the bias source of the SiPM.

Four different kinds of DAQ setups have been used in the measurement.

(A) While measuring the breakdown voltage (V$_{bd}$), the LED is driven by the pulse generator with periodic square pulses of 1~kHz and the oscilloscope is triggered by synchronous pulses. Signals of the LED trigger and SiPM are recorded and the time window is set to 500~ns.

(B) While measuring the relative quantum efficiency (QE), the setups are basically the same as in case A except that the time window is set to 5~$\mu$s and signals from the silicon photocell are also recorded.

(C) While measuring the correlated signal probabilities, the LED is off. Only the SiPM signals are recorded and the oscilloscope is in self-trigger mode. The time window is set to 500~$\mu$s because we aim to analyze the correlated signals which follow the initial.

(D) While measuring DCR, the signals are directly sent to a LTD (CAEN, N844). The output of the LTD is sent to a scaler (CAEN, N145) for counting.

All the signals are recorded by the oscilloscope and the data sampling rate is set to 500~MS/s.

\begin{figure}[htb]
\centering
\includegraphics[height=4.5cm]{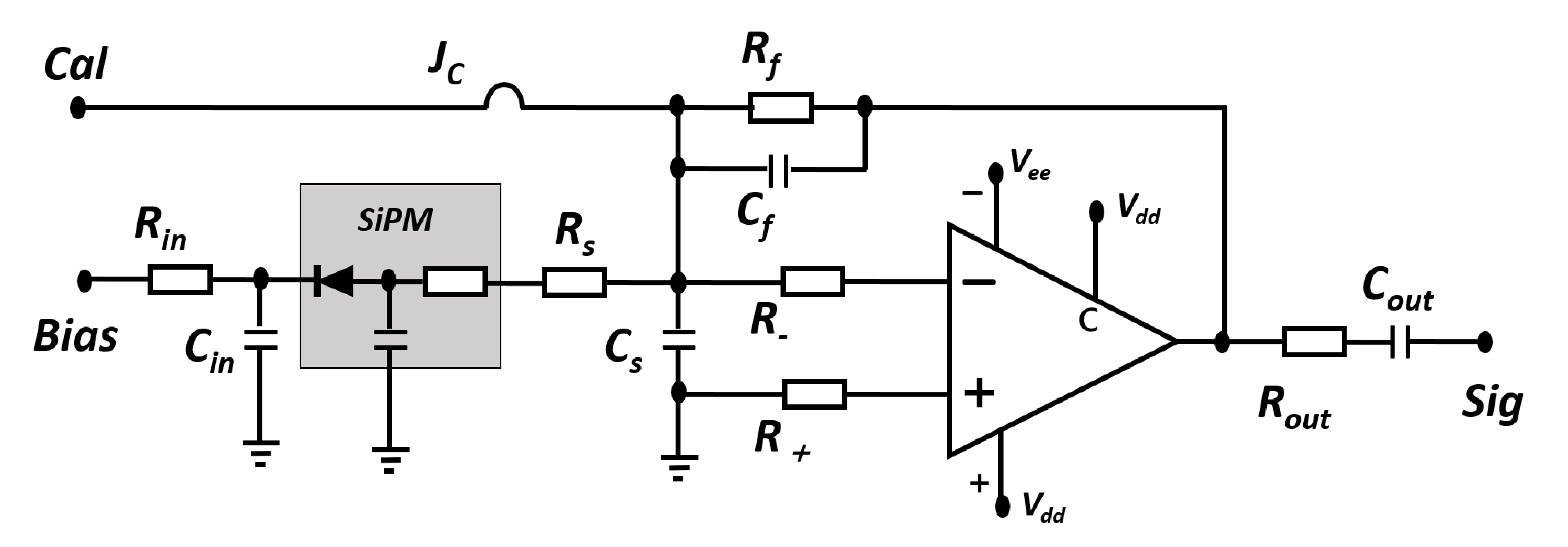}
\caption{Circuit diagram of the readout. Cal is the calibration signal which is generated by the pulse generator. Junction J$_C$ is connected while testing the amplifier stability, otherwise it is disconnected.}
\label{fig.circuit}
\end{figure}

LMH6629 is a high-speed, ultra-low noise amplifier designed for the applications requiring wide bandwidth with high gain and low noise~\cite{LMH6629}. Although the data sheet claims that the amplifier works at 233~K to 398~K, the application reported in Ref.~\cite{DS} shows that it can even work at 60~K. Therefore, in this work, LMH6629 is chosen to amplify the SiPM signals. The circuit diagram inherited from Ref.~\cite{DS}, is shown in Fig.~\ref{fig.circuit} and the values of the components used in this work is presented in Tab.~\ref{tab:value}.

\begin{table}[htb]
\begin{center}
\caption{The values of the components shown in fig.~\ref{fig.circuit} used in this work. }
\begin{tabular}[c]{c| c | c | c | c |c |c} \hline
 Resistance & R$_{in}$ & R$_s$ & R$_f$ &R$_{out}$ & R$_{-}$ & R$_{+}$  \\\hline
 Value & 10~k$\Omega$ & 10~$\Omega$ & 2~k$\Omega$ & 50~$\Omega$ & 2~k$\Omega$ & 5~$\Omega$ \\\hline
 capacitance & C$_{in}$ & C$_s$ & C$_f$ & C$_{out}$ &  &   \\\hline
 Value  & 100~nf & 30~nf & 10~nf & 10~nf & & \\\hline
 \end{tabular}
\label{tab:value}
\end{center}
\end{table}

\section{Stability test of LMH6629}
The tests reported in Ref.~\cite{DS} show that the amplifier LMH6629 can work at 60~K, but its cryogenic performance, especially the gain stability around 87~K, is not clear yet. Thus, before measuring the characterisation of the SiPM, the stability of LMH6629 amplifier needs to be estimated to guarantee the linearity of the SiPM outputs.

\begin{figure}[htb]
\centering
\includegraphics[height=4.5cm]{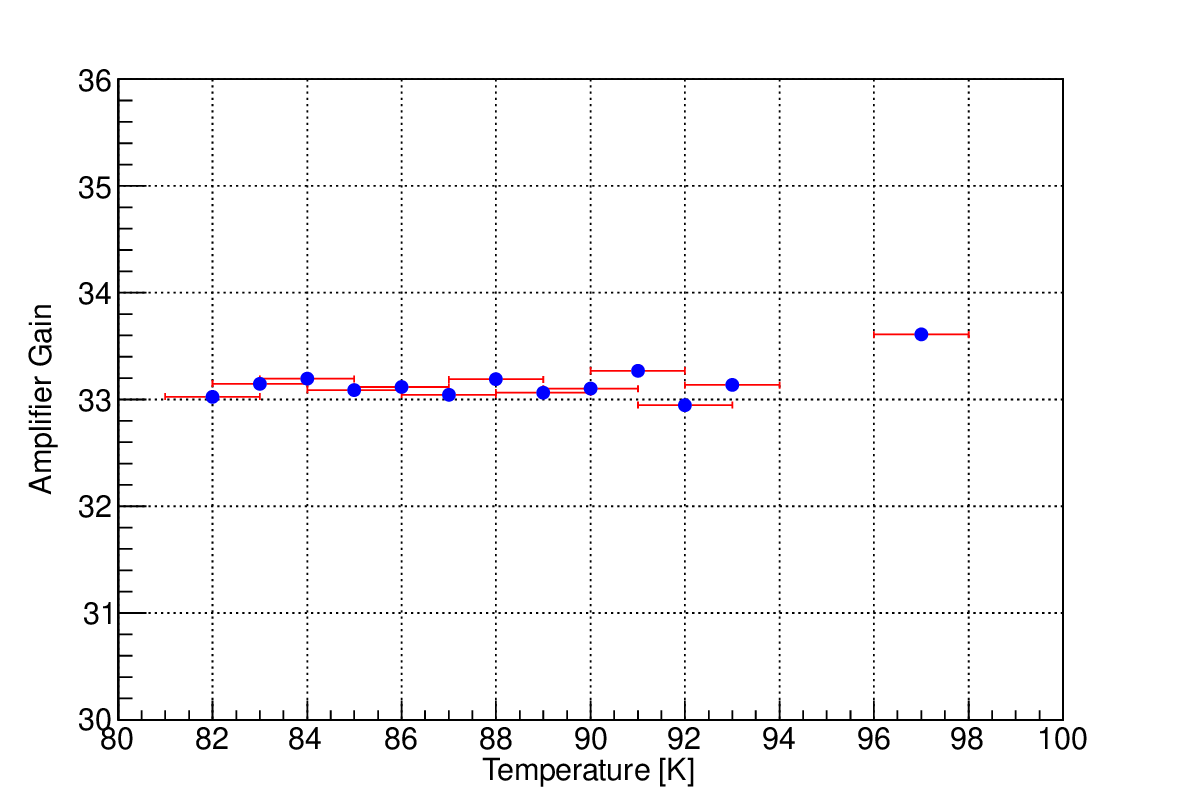}
\caption{Amplifier gain vs. temperature. Gain is calculated by the charge ratio of signals before and after amplification. The errors of the X-axis are the accuracy of the temperature control system and the errors of the Y-axis are statistical only.  }
\label{fig.t_gain}
\end{figure}

The circuit diagram shown in Fig.~\ref{fig.circuit} is used in this measurement, but the SiPM is removed and junction J$_c$ is connected. The input signals, which are periodic square pulses with 1~kHz frequency, are generated by the pulse generator mentioned in Sec.~\ref{sec.setup}. The width of the input pulse is 100~ns and the amplitude is 2~mV. Two identical and synchronised signals are generated by a AFG31102 pulse generator from two output channels. One is directly sent to the oscilloscope for pulses recording, while the other is amplified before being sent to the oscilloscope. Ten thousands events were recorded in each temperature to calculate the charges of the input and output pulses. The charge ratio of the two signals is used to calculate the gain of the amplifier. Fig.~\ref{fig.t_gain} shows the gain variation with temperatures and the stability is within 2\%, which is calculated according to (Maximum-Minimum)/Average. Systematic uncertainties from the temperature controller have been presented in the X-axis.

\begin{figure}[htb]
\centering
\includegraphics[height=4.5cm]{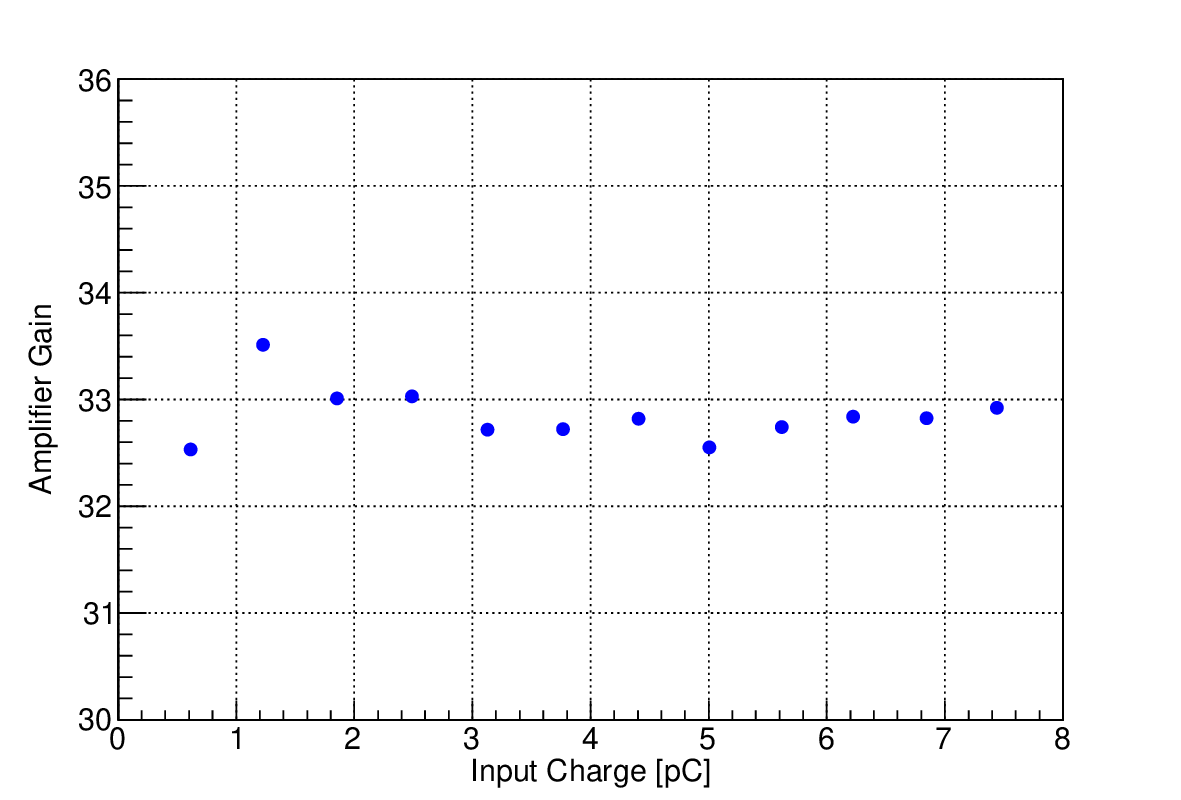}
\caption{Amplifier gain vs. input charge. The errors of X-axis are statistical only while the errors of Y-axis include the statistical and systematic errors, the systematic errors are derived from $\pm$1~K temperature stability. They are too small to see.}
\label{fig.gain}
\end{figure}

Besides temperatures, the input charge may also affect the amplifier gain. According to Ref.~\cite{datasheet}, the typical gain of S13370-6050CN SiPM is 5.8~$\times$10$^6$, thus one photoelectron (P.E.) corresponds to a charge of $\sim$0.9~pC. In the test of gain variation with the input charge, the width of the input pulse is set to 100~ns which is the typical width of the SiPM output signal, while the amplitude changes from 2~mV to 15~mV which is equivalent to the input charge from 0.5~pC to 7.5~pC. The result is shown in Fig.~\ref{fig.gain} and the variation is within 3\%.

\section{Data Analysis}
\subsection{Signal composition}
\label{sec.sc}

\begin{figure}[htb]
\centering
\includegraphics[height=4.5cm]{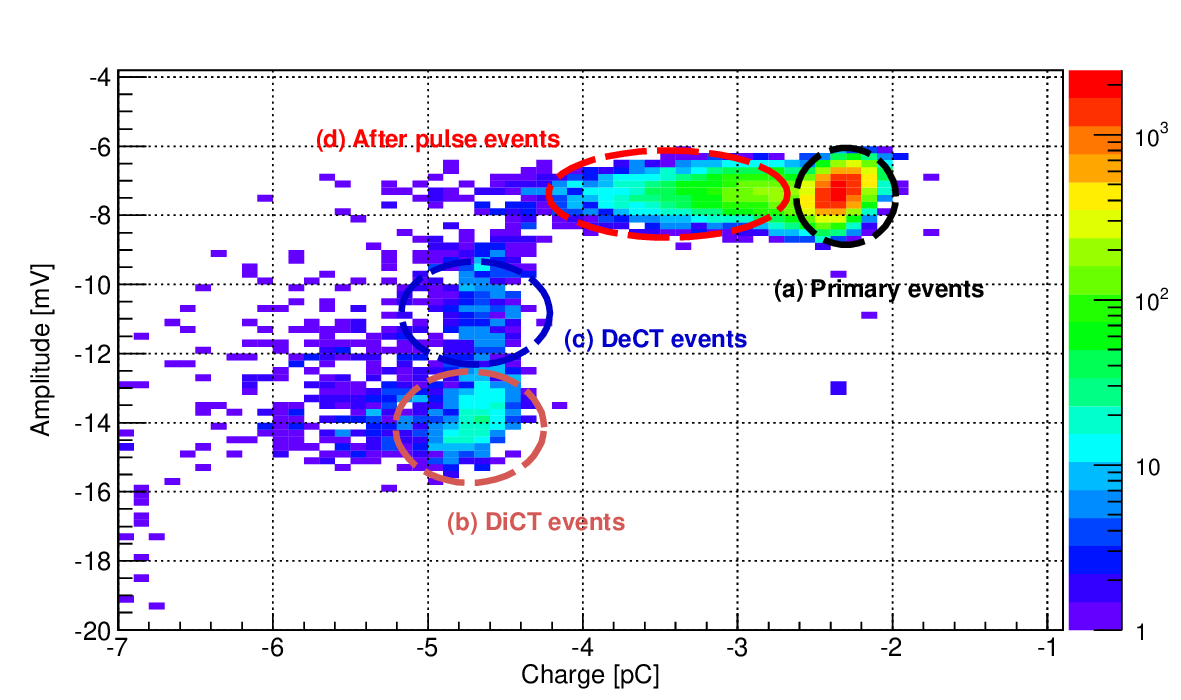}
\caption{Scatter distribution of charge vs. amplitude of S13370-6050CN SiPM operated at 87~K with an over voltage of 3~V. Amplitude and charge are negative because the pulses are negative. Events in the black circle are dark noises, in the red circle are after pulses, in the blue circle are delayed crosstalks and in the orange circle are direct crosstalks. }
\label{fig.A_C}
\end{figure}

\begin{figure}[htb]
\centering
\includegraphics[width=6.5cm,height=4cm]{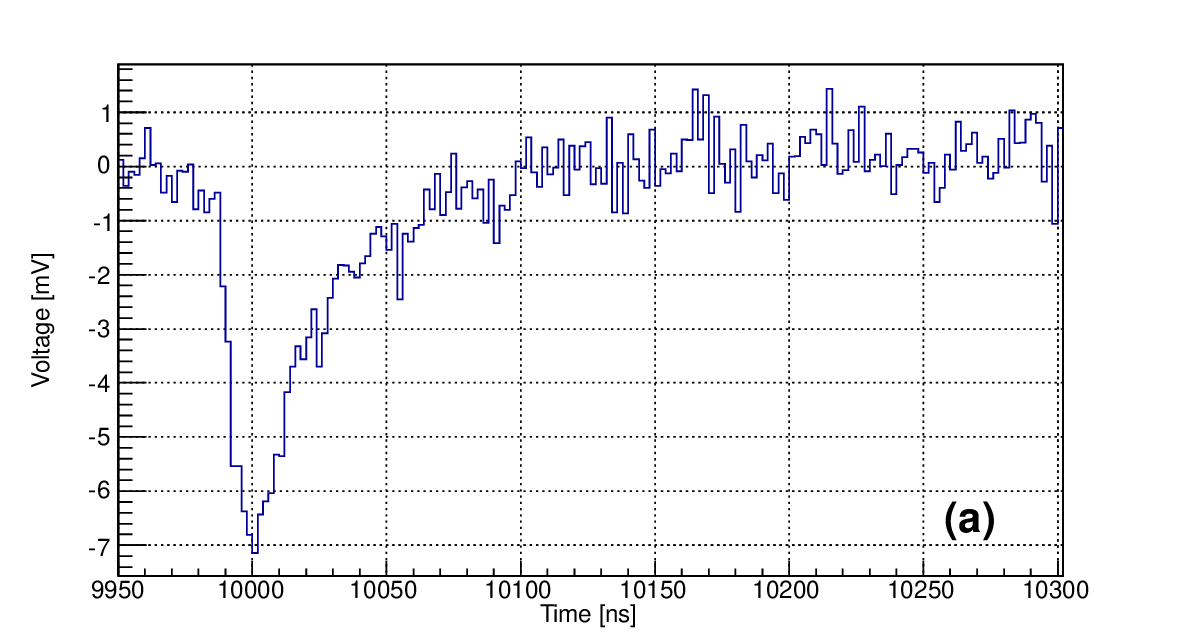}
\includegraphics[width=6.5cm,height=4cm]{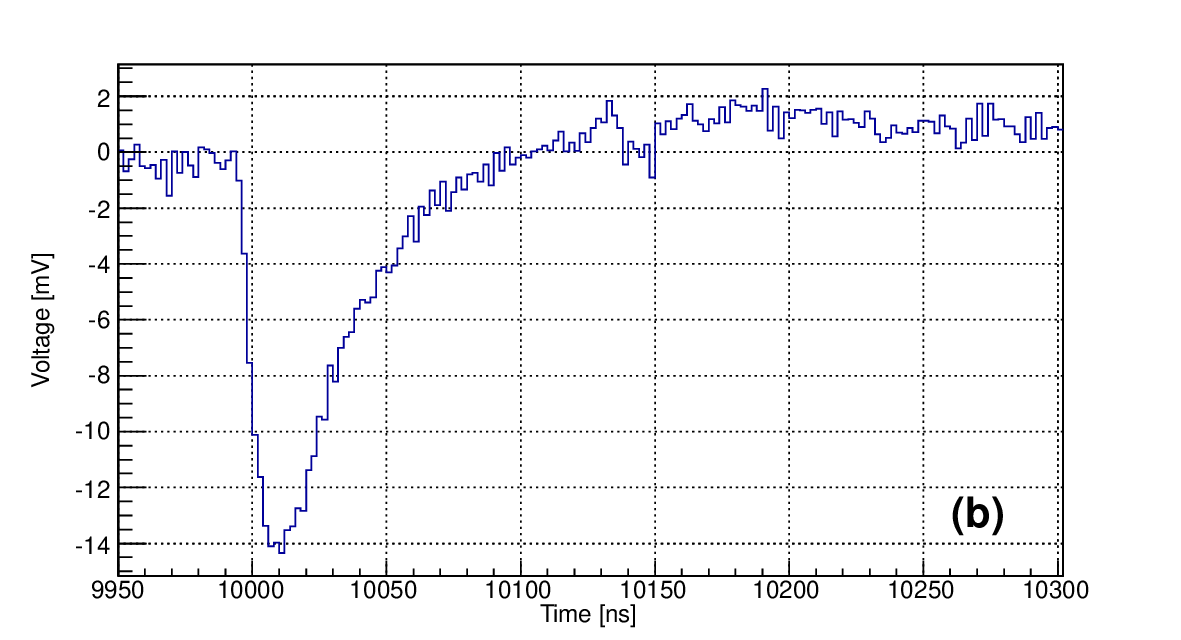}
\includegraphics[width=6.5cm,height=4cm]{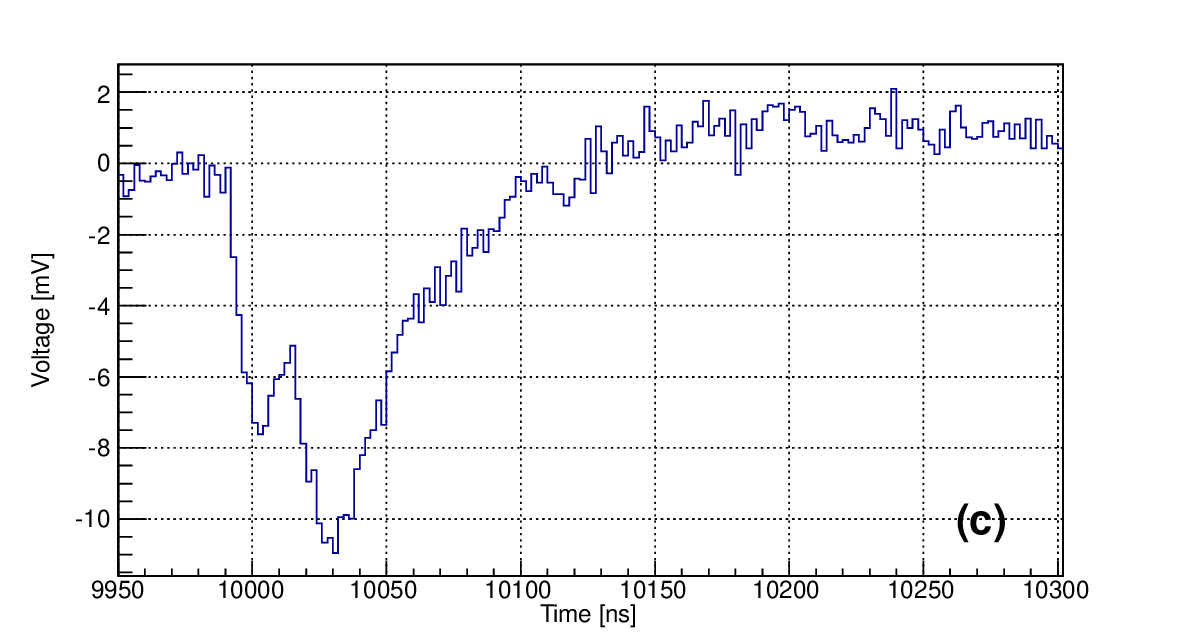}
\includegraphics[width=6.5cm,height=4cm]{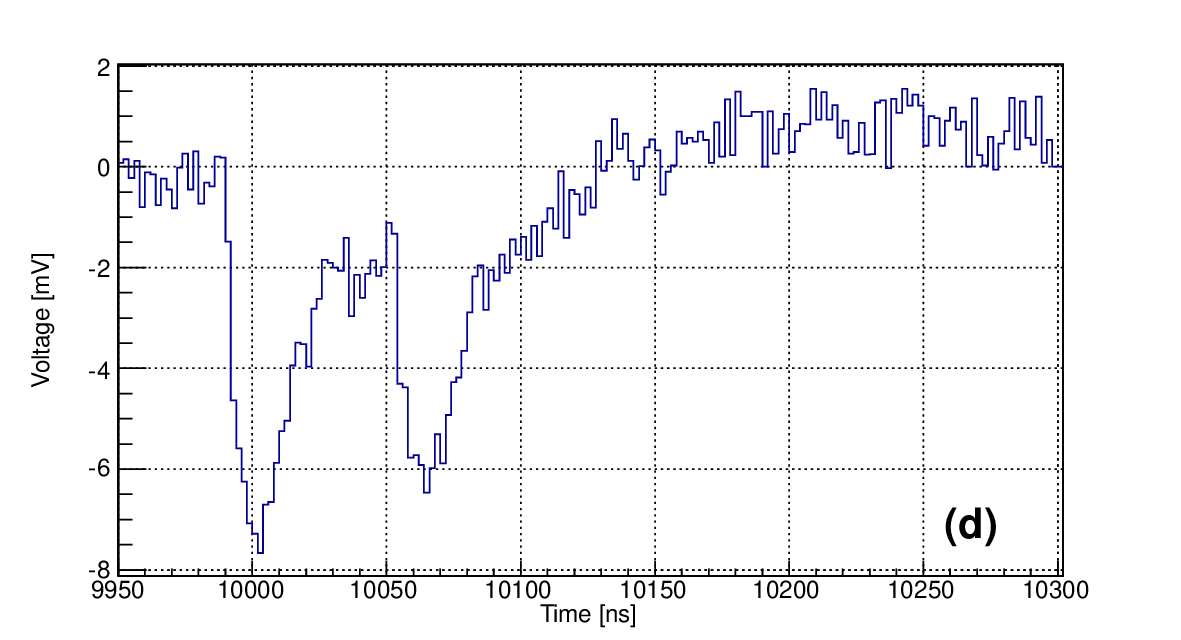}
\caption{Example pulses for different events. (a) SPE event. (b) Direct crosstalk (DiCT). (c) Delay crosstalk (DeCT). (d) After pulse (AP).}
\label{fig.example}
\end{figure}

A custom data analysis program has been developed and a detailed analysis has been performed. For each event, the program calculates the amplitude and charge of the waveform. For simplicity, the amplitude of single photoelectron (SPE) is normalized to 1~P.E.A. and the charge of single photoelectron is normalized to 1~P.E.C.. Baseline has been subtracted while calculating the amplitude and charge. An example of a scatter distribution of charge and amplitude is shown in Fig.~\ref{fig.A_C}. The SiPM is operated at 87~K with 3~V over voltage (V$_{over}$), which is the difference between the supplying bias and the breakdown voltage. The events shown in Fig.~\ref{fig.A_C} consist of the following four parts~\cite{sensl,DSsipm,NIMA620}:

A. Dark noise : In the absence of light, dark noise, which is produced by carriers generated in the depletion region, constitutes the main part of the signal. The dark noise is SPE event thus has an amplitude centred at 1~P.E.A. and a charge centred at 1~P.E.C.. A sample pulse of dark noise is shown in Fig.~\ref{fig.example}-(a).

B. Direct crosstalk: Direct crosstalk (DiCT) is caused by a photon which is emitted by the movement of accelerated carriers in the strong field during the primary avalanche triggering a neighbouring pixel within an extremely short time interval. Thus, a DiCT signal has an amplitude centred at 2~P.E.A. and a charge centred at 2~P.E.C., which is just similar to a 2~P.E. event. A sample pulse of DiCT is shown in Fig.~\ref{fig.example}-(b).

C. Delayed crosstalk: Delayed crosstalk (DeCT) occurs when the photons produced in the primary avalanche absorbed in the non-depleted region of a neighbouring cell. Before triggering a second avalanche, the carriers have to diffuse into the high-field region, thus the time difference between the two avalanches is around several nanosecond. Based on its mechanism, the amplitude of DeCT should be between 1~P.E.A. to 2~P.E.A. while the charge should be centred at 2~P.E.C.. An example pulse of DeCT is shown in Fig.~\ref{fig.example}-(c).

D. After pulse: After pulse (AP) is observed when a secondary electron is trapped by some sort of impurity during the primary avalanche. The trapped electron is then released after a characteristic time from nanosecond to microsecond and finally results a second avalanche. In principle, the time interval and energy distribution depend on the location of the trapped electron. The AP events should have an amplitude around 1~P.E.A. while its charge should be between 1~P.E.C. to 2~P.E.C. because the second avalanche discharge starts in the middle. The example pulse of AP is shown in Fig.~\ref{fig.example}-(d).

\subsection{Pulse fitting}
\label{sec.pulsefitting}
As described in Sec.~\ref{sec.sc}, different kinds of signals can de identified with the charge and amplitude information. In order to better estimate the charge and time information of DeCT and AP events, a pulse fitting program has been developed.

\begin{figure}[htb]
\centering
\includegraphics[height=5.5cm]{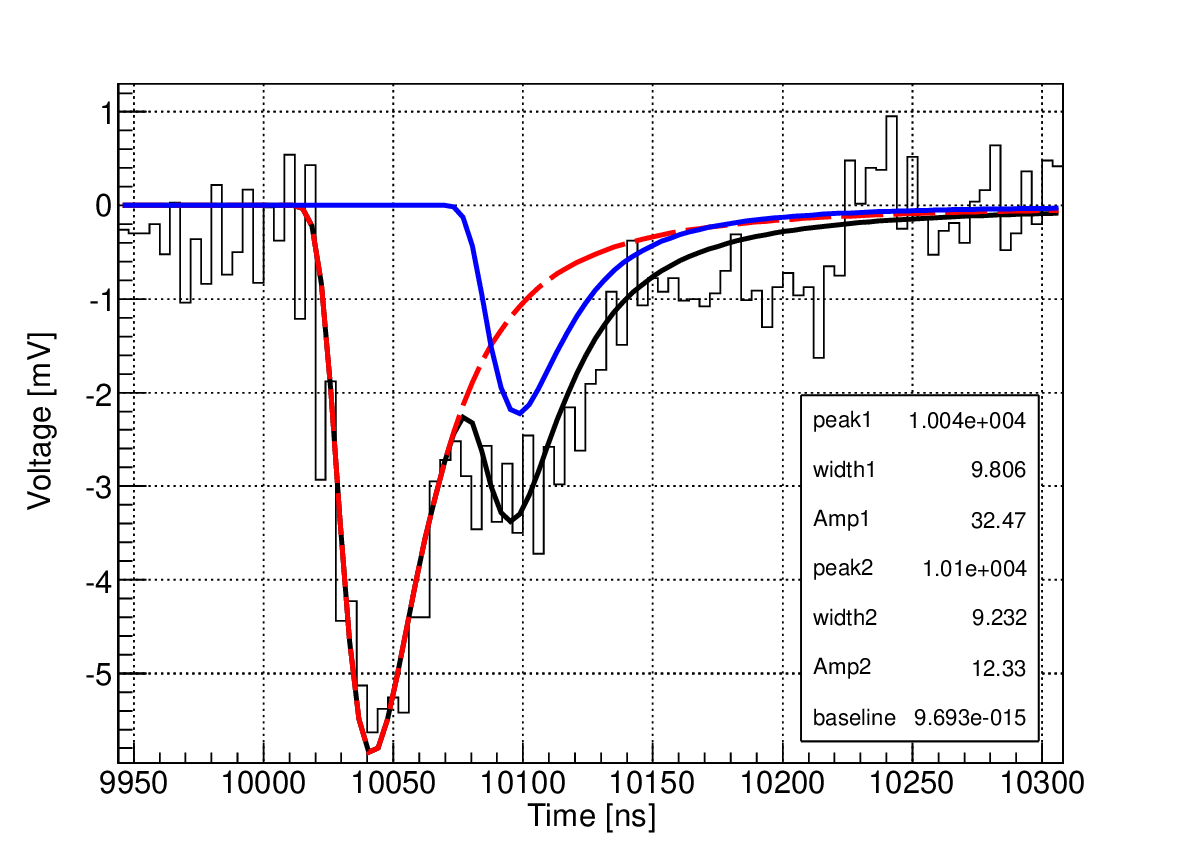}
\caption{A sample of AP event fitted according to Eq.~\ref{Eq.landau}. The black curve is the total waveform, the red dashed curve is the primary pulse, the blue curve is the after pulse. Baseline is not plotted because it is very small.}
\label{fig.fit}
\end{figure}

As can be seen from Fig.~\ref{fig.example}-(c) and Fig.~\ref{fig.example}-(d), the delayed pulse overlaps on the primary SPE pulse, thus the total waveform can be described according to Eq.~\ref{Eq.landau}, which is a superposition of two Landau Distributions~\cite{Landau} and a constant with seven parameters.
\begin{equation}
Fit(x)=A_1Landau(x,p_1,w_1)+A_2Landau(x,p_2,w_2)+BL
\label{Eq.landau}
\end{equation}

where A$_1$, A$_2$ represent the amplitudes, p$_1$, p$_2$ represent the locations and w$_1$, w$_2$ represent the widths of the two pulses, respectively, BL represents the baseline. The first term of Eq.~\ref{Eq.landau} is for describing the primary pulse and the second term is for the delayed pulse. Fig.~\ref{fig.fit} shows an example of the fitting result of an AP event. DeCT event pulses have the same shape as AP events, thus Eq.~\ref{Eq.landau} can be used to fit both kinds of signals.

\section{Results}

DCR and total correlated noise probability changes with the over voltage, therefore the breakdown voltage must be accurately measured.

\subsection{Breakdown voltage}
The breakdown voltage (V$_{bd}$) is the bias point at which the electric field strength generated in the depletion region is just sufficient to produce a Geiger discharge. When the bias is V$_{bd}$, the gain of a SiPM, which is defined as in Eq.~\ref{Eq.gain}, drops to exactly zero.

\begin{equation}
G = \frac{Q}{e}=\frac{\int I(t)dt}{q_e}=\frac{\int V_{out}dt}{R_{load}G_{Amp}q_e}
\label{Eq.gain}
\end{equation}

where G is the gain of the SiPM, Q is the charge of one avalanche, q$_e$ is the charge of an electron, I is the current, V$_{out}$ is the output amplitude of the signal, R$_{load}$ is the matching resistance of the oscilloscope, which is 50~$\Omega$, G$_{Amp}$ is the gain of the preamplifier, which is around 33 in this test.

\begin{figure}[htb]
\centering
\includegraphics[height=4.5cm]{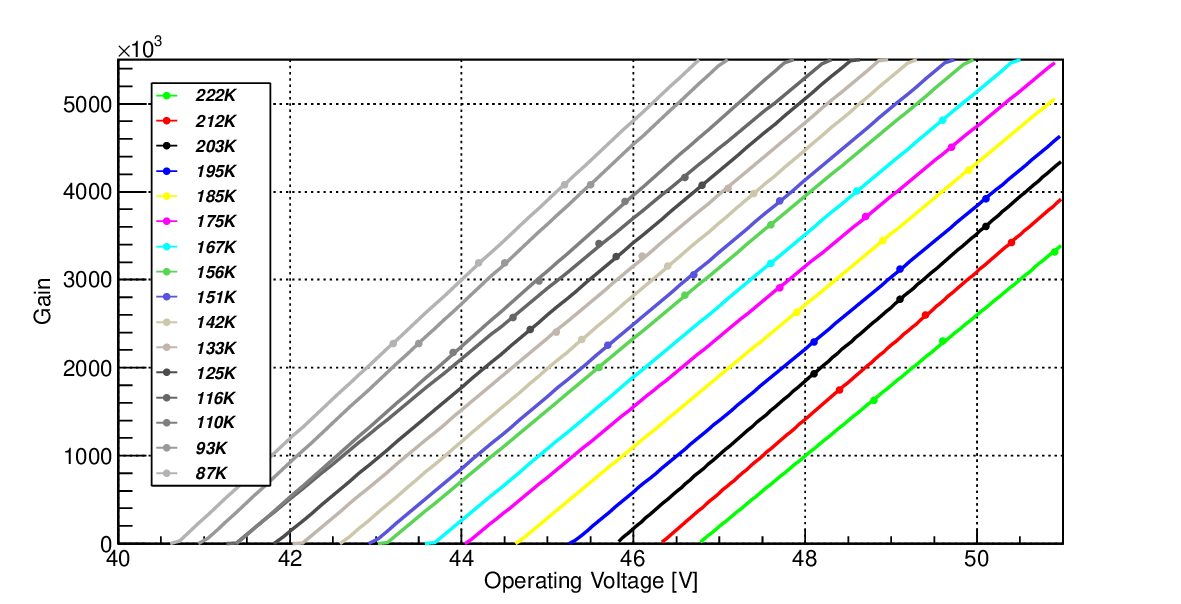}
\caption{Gain curves at different temperature, errors in the X-axis is the accuracy of the bias supplier, which is 1~mV, errors in the Y axis are statistical only. }
\label{fig.vbd1}
\end{figure}

\begin{figure}[htb]
\centering
\includegraphics[height=4.5cm]{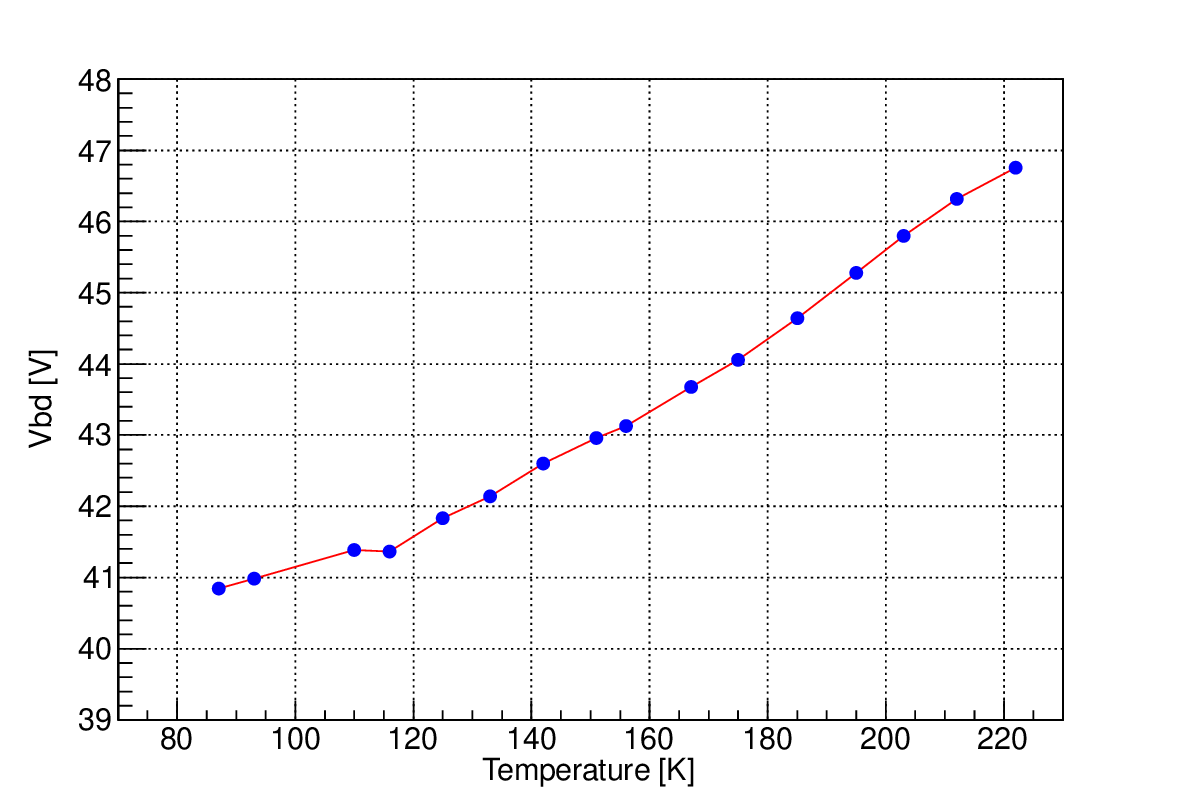}
\caption{Change of V$_{bd}$ with temperature, the errors in the X-axis are the accuracy of the temperature system, which is $\pm$1~K, the errors in the Y-axis are statistical only but too small to see.}
\label{fig.vbd2}
\end{figure}

V$_{bd}$ can be calculated as the intercept of linear fits with X-axis which is presented in Fig.~\ref{fig.vbd1}. Fig.~\ref{fig.vbd2} shows the change of V$_{bd}$ with temperature. Above 120~K, V$_{bd}$ decreases linearly with temperature, which is about 0.5~V/10~K, while below 120~K the decrease tends to be slow.

\subsection{Dark counting rate (DCR)}
\begin{figure}[htb]
\centering
\includegraphics[height=4.5cm]{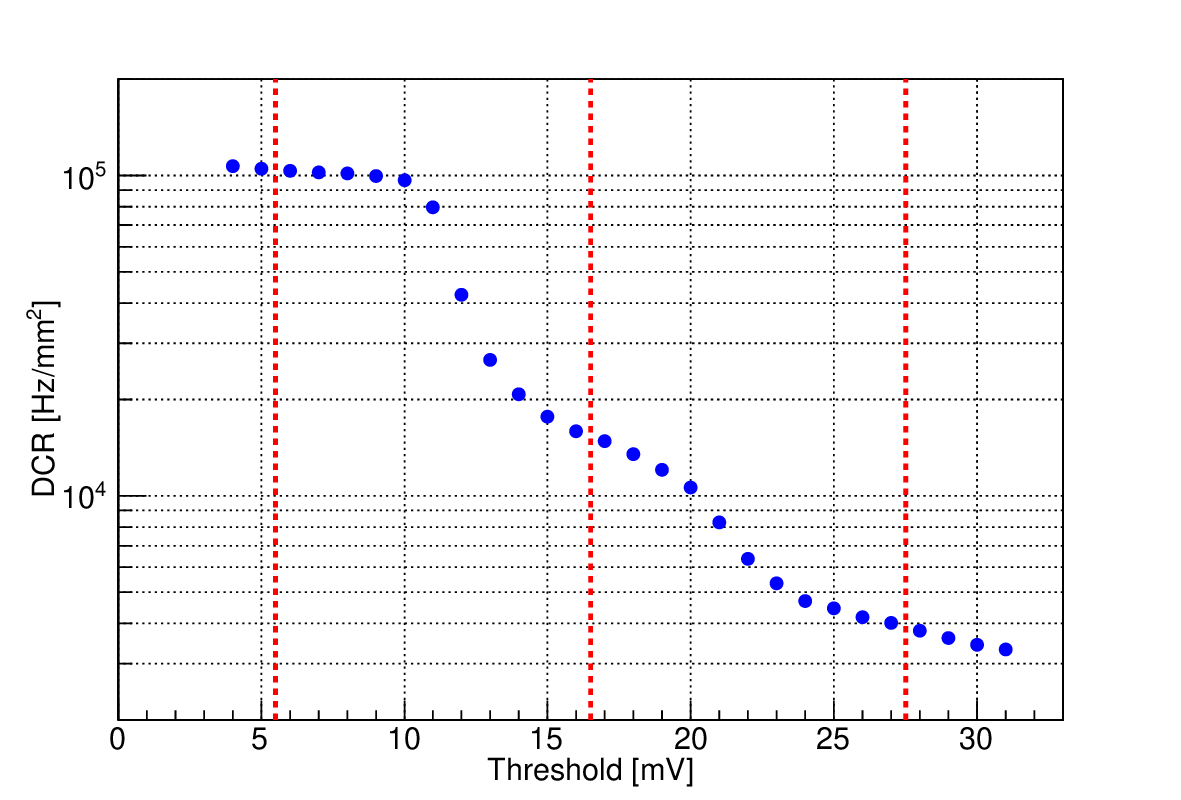}
\caption{DCR of the SiPM operated at V$_{over}$=4~V at 298~K as a function of the discriminator threshold. The locations of 0.5~P.E., 1.5~P.E., and 2.5~P.E. are indicated in the plot by vertical dotted lines separately.}
\label{fig.DCR_1}
\end{figure}

The dark current counting rate is due to thermally generated electrons that trigger an avalanche in the active volume. DCR can be measured with a counting system with a threshold of 0.5~P.E. level. The change of DCR with threshold is shown in Fig.~\ref{fig.DCR_1} and the event rates at 0.5~P.E., 1.5~P.E., and 2.5~P.E. are indicated in the plot by the vertical dotted lines.

\begin{figure}[htb]
\centering
\includegraphics[width=6cm]{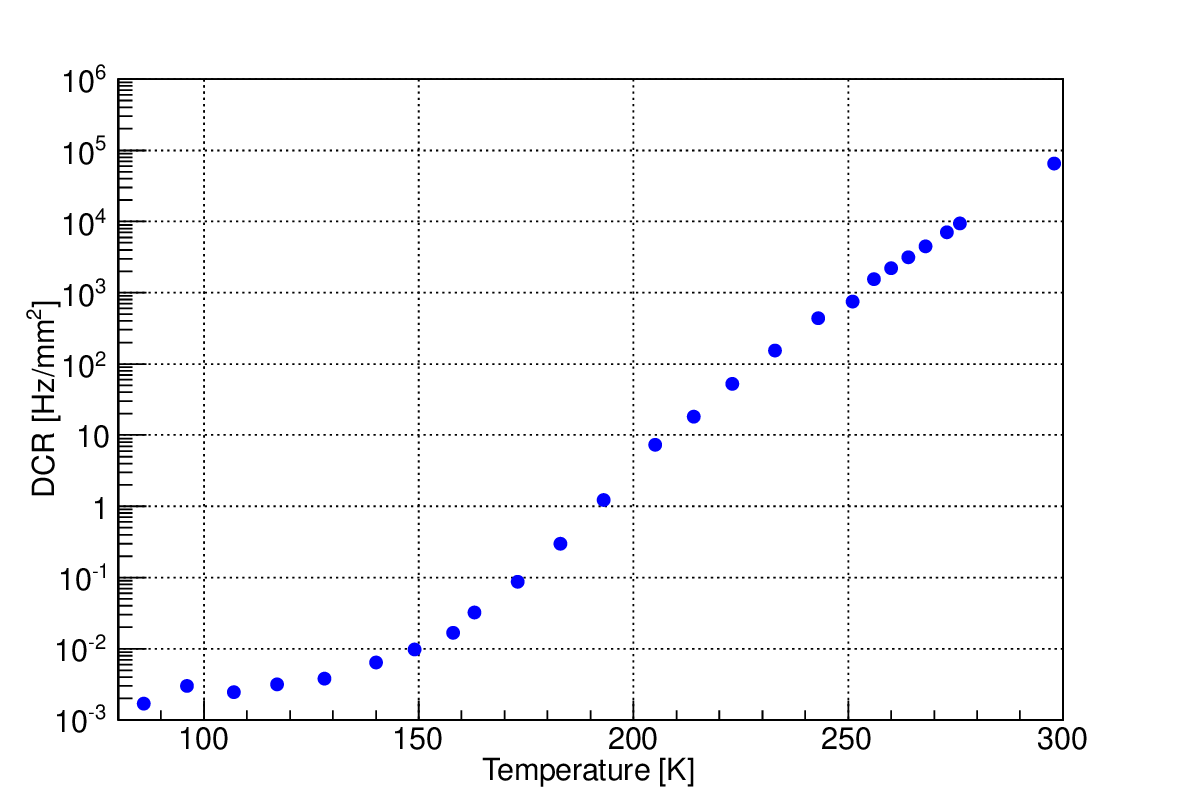}
\includegraphics[width=6cm]{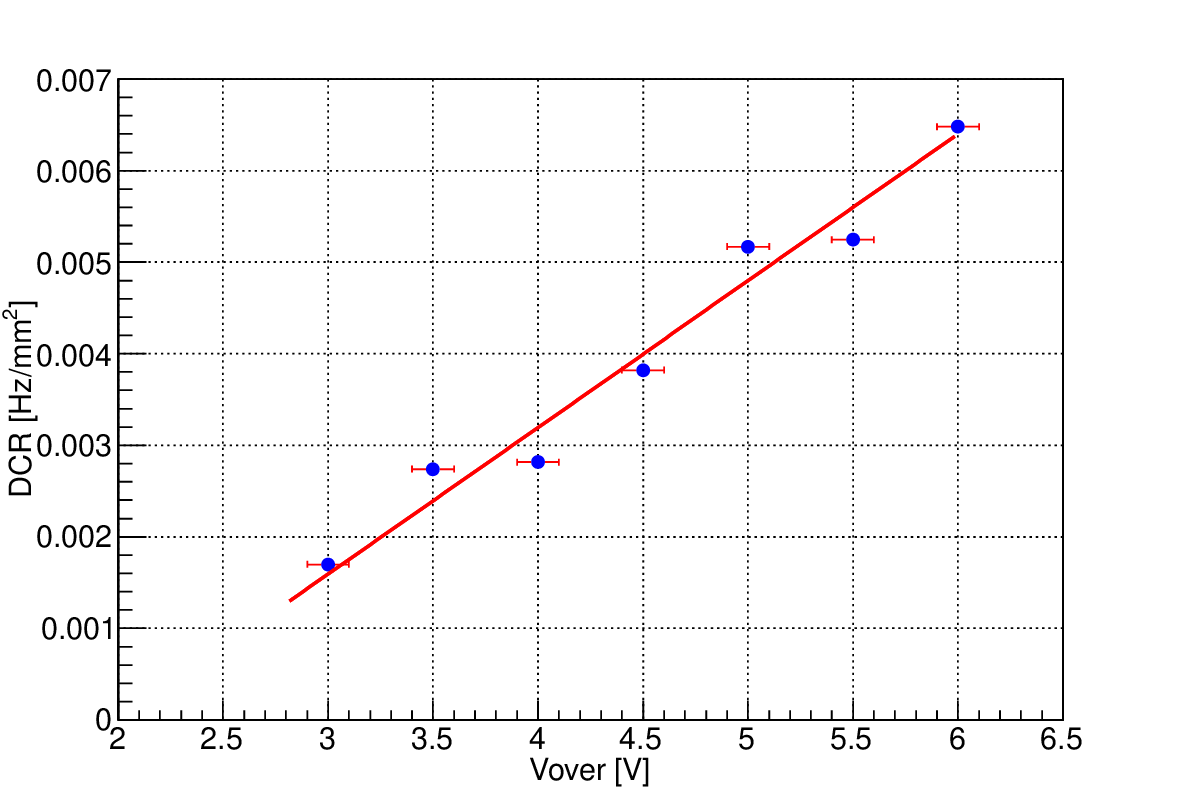}
\caption{Left: Change of DCR with temperatures, V$_{over}$ is set to 4~V during the test, errors of the X-axis are the temperature accuracy of $\pm$1~K, errors of the Y-axis are statistical only. Right: Change of DCR with V$_{over}$ at 87~K, the errors shown in the X-axis are caused by the temperature variation and errors in the Y-axis are statistical only.}
\label{fig.DCR_2}
\end{figure}

As discussed in Ref.~\cite{sensl,NIMA620,DSsipm}, DCR depends on both temperature and V$_{over}$. The study of those dependences, shown in Fig.~\ref{fig.DCR_2}, indicate that the DCR of the tested SiPM is $\sim$0.003~Hz/mm$^2$ at 87~K with an over voltage of 4~V.

\subsection{Correlated signals}
\begin{figure}[htb]
\centering
\includegraphics[height=4.5cm]{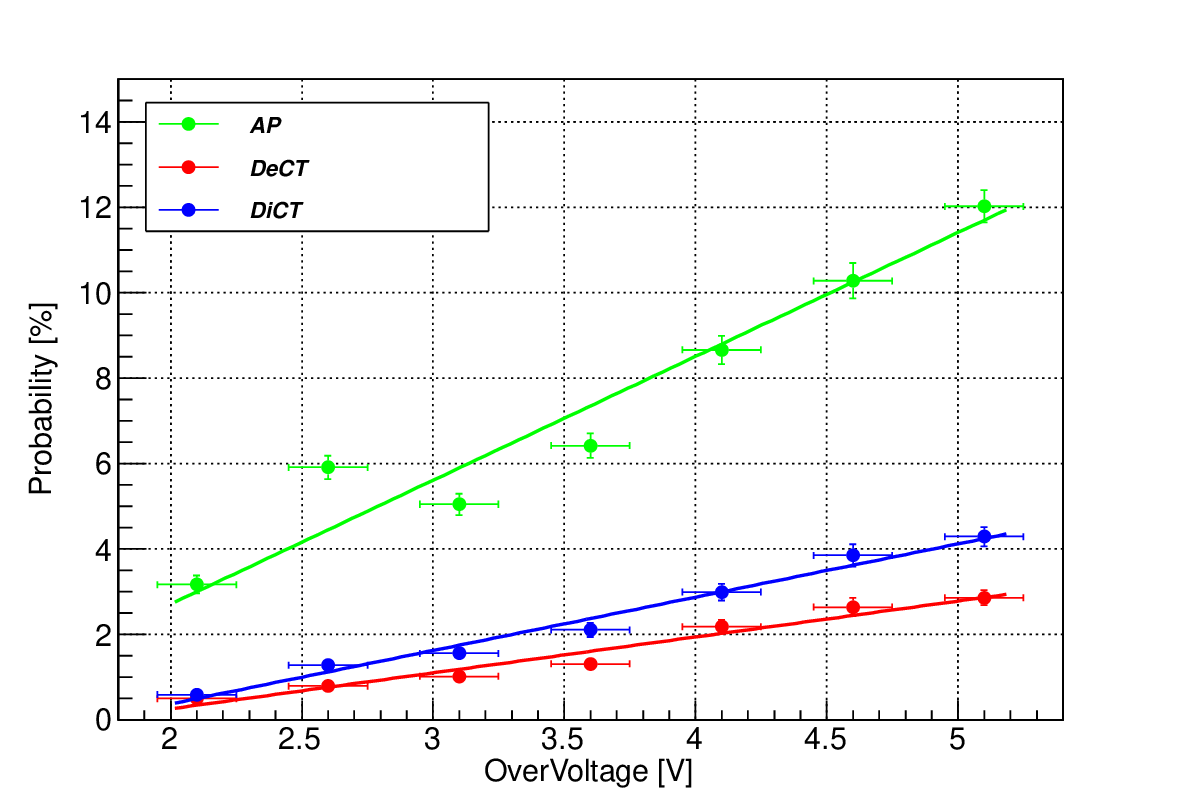}
\caption{The changes of the probabilities of DiCT, DeCT and AP with V$_{over}$ at 87K. The errors in the X-axis are caused by the temperature variation and the errors in the Y-axis is statistical only.}
\label{fig.relatedsignals}
\end{figure}

Correlated signals are the general term for DiCT, DeCT and AP, the definitions and characteristics of which are introduced in Sec.~\ref{sec.sc}. Fig.~\ref{fig.relatedsignals} shows the changes of the probabilities of DiCT, DeCT and AP with different V$_{over}$. At 87K, the total correlated signal probability is $\sim$10\% at V$_{over}$~=~3~V, which could satisfy the requirements of using SiPM in a dual phase LAr detector~\cite{DSyellowbook}.

\begin{figure}[htb]
\centering
\includegraphics[height=4.5cm]{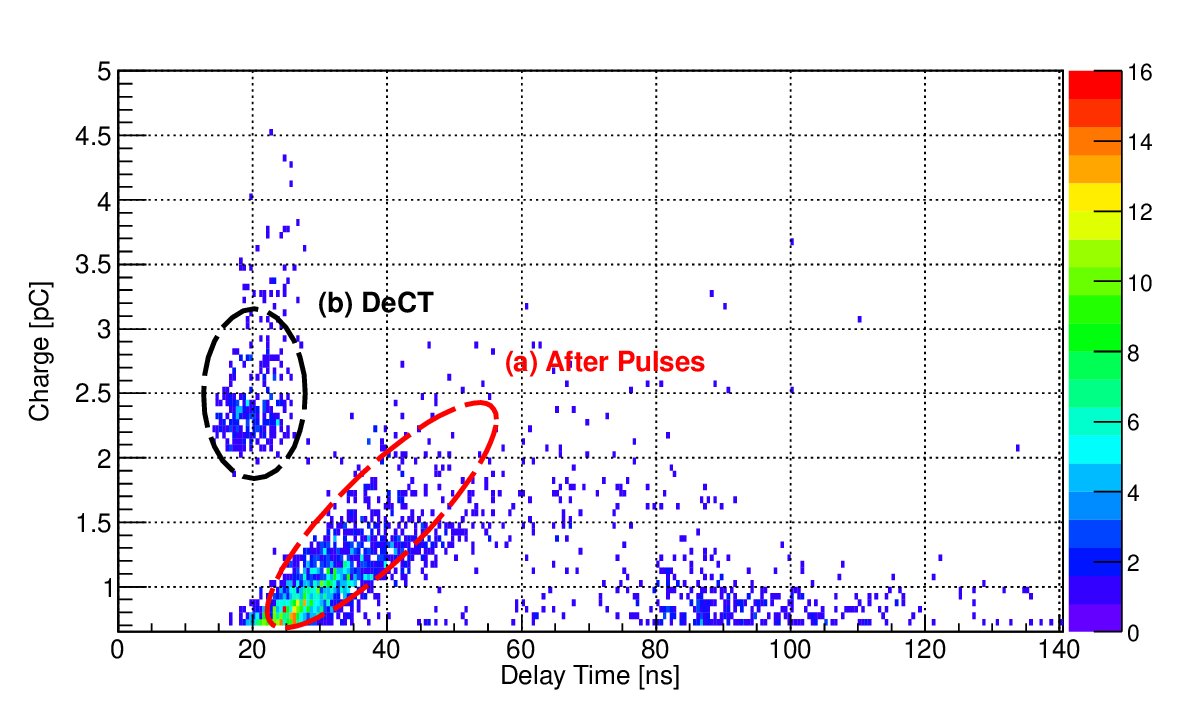}
\caption{The time difference between correlated signals and primary DCR vs. the charge of the correlated signals (DeCT and AP). Here, the charge is shown with its absolute value, without sign. X-axis is the time difference between the delayed pulse and the primary pulse, Y-axis is the energy of the delayed pulse. Events in the red circle are APs and in the black circle are DeCTs. The events in the time range of 80-100~ns with charge less than 1~pC are cases where the pulse shape first failed. They are regular DCR pulses with somewhat larger baseline jitter.}
\label{fig.scatterplots_time_charge}
\end{figure}

The time interval distribution between correlated pulses (DeCT and AP) and the primary DCR is shown in Fig.~\ref{fig.scatterplots_time_charge}. X-axis is the time difference between the delayed pulse (p$_2$) and the primary pulse(p$_1$), which is calculated as p$_2$-p$_1$ of Eq.~\ref{Eq.landau}. Y-axis is the charge of the delayed pulse, which is the integral of the blue line shown in Fig.~\ref{fig.fit}. The plot shows that there is a direct proportion between the delay time and the pulse energy for AP events, which possibly indicates that the further the electron is trapped from the anode (higher charge), the faster the electron is released from the impurity (shorter time interval).

\subsection{Relative quantum efficiency}
\begin{figure}[htb]
\centering
\includegraphics[height=4.5cm]{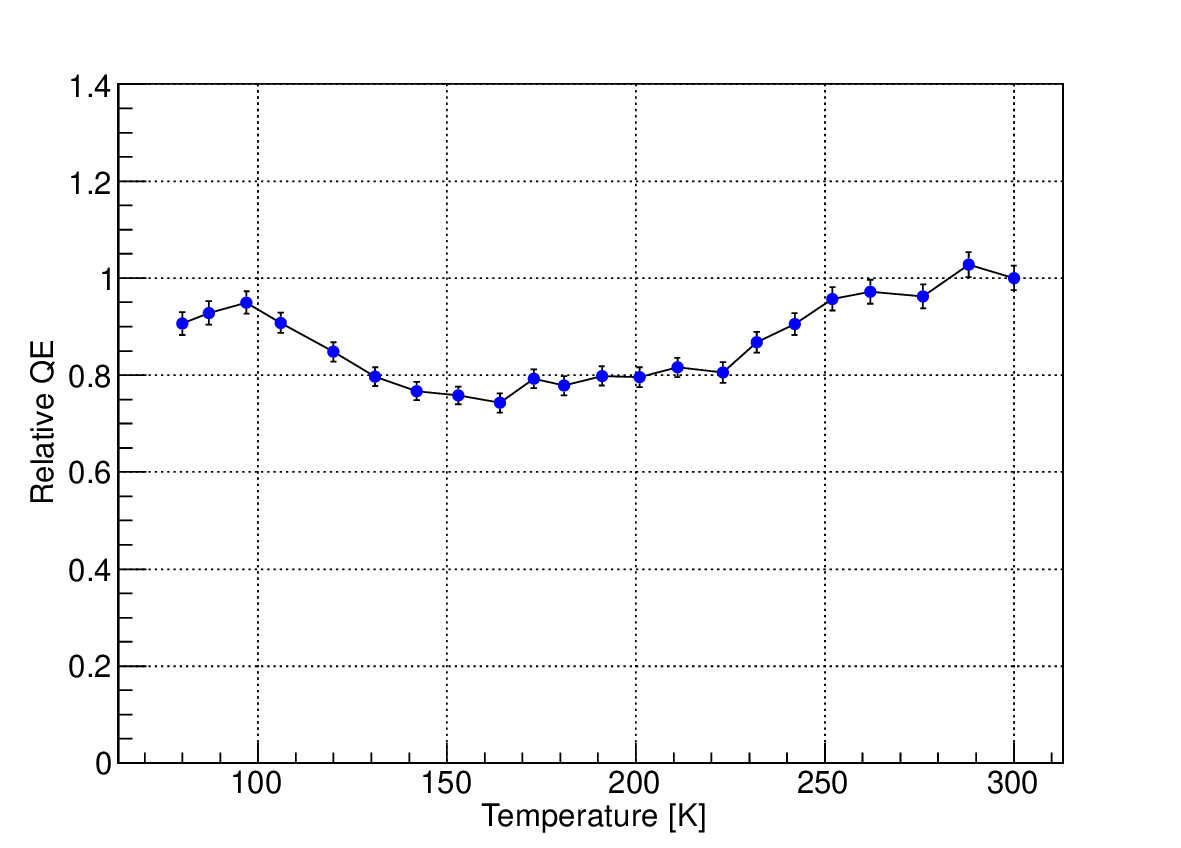}
\caption{The relative quantum efficiency at different temperatures. The results are normalized to the point of 300~K. Errors in the X-axis are the temperature accuracy of $\pm$1~K and the errors in the Y-axis are statistical only.}
\label{fig.qe}
\end{figure}

One of the advantages of S13370-6050CN SiPM is that it can directly detect the photons emitted by liquid argon, of which the wavelength is peaked at 128~nm. But the data sheet also claims that it has the largest quantum efficiency at $\sim$ 500~nm~\cite{datasheet}. For liquid argon detectors, some~\cite{DS50,DEAP} prefer not to read the 128~nm photons directly, one reason is that the photon sensors have very low quantum efficiency at 128~nm and another reason is that the wavelength shifter, 1,1,4,  4-tetraphenyl-1,3-butadiene (TPB)~\cite{TPB}, has more than 100\% photon conversion efficiency~\cite{NIMA654}, which means that more than one 420~nm photons will be emitted when one 128~nm photon is absorbed. According to our current experimental scheme, TPB will also be used. In Ref.~\cite{datasheet}, the quantum efficiency is measured at 298~K, while according to Ref.~\cite{QE}, the quantum efficiency  at room temperatures and liquid argon temperature may be very different. In this work, a 425~nm LED is used to measure the relative quantum efficiency with an over voltage of 3~V. The result, plotted in  Fig.~\ref{fig.qe}, indicates that the quantum efficiency difference of S13370-6050CN SiPM between 300~K and 87~K is $\sim$10\%.

\section{Summary}
A program of using liquid argon to detect $\bar{\nu_e}$-Ar CE$\nu$NS at Taishan Nuclear Power Plant has been proposed and SiPM will be used as the photon sensors. In order to study the possibility of using S13370-6050CN SiPM in our LAr detector, a cryogenic system has been developed and the temperature dependences of V$_{bd}$, DCR, correlated noises and relative QE has been measured in detail. Our results show that the DCR is below 0.003~Hz/mm$^2$ and the total correlated noise is less than 10\% at 87~K with 3~V over voltage, which indicate that the S13370-6050CN SiPM made by Hamamatsu is a good candidate for our liquid argon detector.

\section{Acknowledgements}
This work is supported by National Key R\&D Program of China (Grant No. 2016YFA0400304) and National Natural Science Foundation of China (Grant No. 12020101004).

The authors would like to thank Alessandro Razeto, Paolo Musico and Gemma Testera of INFN for their help during the design of the readout system.


\end{document}